\documentclass[aps,prl,twocolumn,longbibliography,superscriptaddress,floatfix,10pt]{revtex4-1}

\usepackage{amsfonts,amssymb,dsfont}
\usepackage[sumlimits,intlimits]{amsmath}
\usepackage{graphics}
\usepackage{graphicx}
\usepackage[dvipsnames]{xcolor}
\usepackage{color}
\usepackage{mathrsfs}
\usepackage{textcomp}
\usepackage{verbatim}
\usepackage[colorlinks=true,citecolor=blue]{hyperref}
\usepackage{bm}
\usepackage{soul}
\usepackage{braket}
\usepackage{times}
\usepackage{booktabs}
\usepackage{lipsum}
\usepackage[final]{pdfpages}
\usepackage{pgffor}
\makeatletter
\AtBeginDocument{\let\LS@rot\@undefined}
\makeatother

\usepackage[capitalise,compress]{cleveref}
\crefrangelabelformat{equation}{\textup{(#3#1#4)}--\textup{(#5#2#6)}}

\newcommand{\bS}{\mathbf{S}}

\begin{document}

\title{Non-equilibrium criticality in quench dynamics of infinite-range spin models}

\author{Paraj Titum}
\affiliation{Joint Quantum Institute, NIST/University of Maryland, College Park, Maryland 20742, USA}
\affiliation{Joint Center for Quantum Information and Computer Science, NIST/University of Maryland, College Park, Maryland 20742, USA}
\affiliation{Johns Hopkins University Applied Physics Laboratory, Laurel, Maryland 20723, USA}
\author{Mohammad F. Maghrebi}
\affiliation{Department of Physics and Astronomy, Michigan State University, East Lansing, Michigan 48824, USA}

\begin{abstract}
Long-range interacting spin systems are ubiquitous in physics and exhibit a variety of ground state disorder-to-order phase transitions. We consider a prototype of infinite-range interacting models known as the Lipkin-Meshkov-Glick (LMG) model describing the collective interaction of $N$ spins, and investigate the dynamical properties of fluctuations and correlations after a sudden quench of the Hamiltonian. Specifically, we focus on critical quenches, where the initial state and/or the quench Hamiltonian are critical. 
Depending on the type of quench, we identify three distinct behaviors where both the short-time dynamics and the stationary state at long times are effectively thermal, quantum, and genuinely non-equilibrium, characterized by distinct universality classes and  static and dynamical critical exponents. These behaviors can be identified by an infrared effective temperature that is finite, zero, and infinite (the latter scaling with the system size as $N^{1/3}$), respectively. The quench dynamics is studied through a combination of exact numerics and analytical calculations utilizing the non-equilibrium Keldysh field theory. Our results are amenable to realization in experiments with trapped-ion experiments where long-range interactions naturally arise.
\end{abstract}

\maketitle

The dynamics of isolated quantum systems has intrigued physicists since the dawn of quantum mechanics \cite{von_Neumann29}. Furthermore, this topic has been in the spotlight in the past twenty years thanks to the experimental advances in ultracold atoms~\cite{Bloch_review_2008,Polkovnikov_review_2011,Schmiedmayer_Review2015} and trapped ions~\cite{Blatt2012} among others~\cite{Schauss_2018,Kimble_2018}. These platforms are some of the  prominent candidates for the quantum simulation of quantum phases of matter, but they are also well suited to investigate the dynamics away from equilibrium. A typical 
experimental setting is one where a system parameter suddenly changes -- a scenario commonly described as a quantum quench. 

There is mounting evidence, both theoretical and experimental, that generic non-integrable systems thermalize upon a quantum quench and local correlations are best described by a finite-temperature ensemble~\cite{DAlessio2016,Gogolin2016,Srednicki1999-eth}. On the other hand, integrable systems defined by an extensive set of conserved quantities fail to thermalize and instead are described by generalized Gibbs ensembles that also take into account all conserved quantities~\cite{Vidmar2016}. But, even integrable systems often thermalize in a weaker sense of thermalization if their \textit{long-wavelength} properties are described by a finite effective temperature. For example, such effective thermal behavior has been identified in one-dimensional condensates \cite{Lamacraft07,Bistritzer07,Kitagawa11,Agarwal14,Agarwal17, Altman15review} and even observed in experiments \cite{Gring12, Smith13}; similar behavior is predicted in integrable $O(N\rightarrow\infty)$ models \cite{Chandran13, Gambassi-Mitra15, Smacchia15,Mitra-Gambassi16,Mitra-Review18}. This weaker notion of thermalization (with obvious merits for critical properties) is one that we adopt in this work.
A natural question is then if, upon a quantum quench and depending on the initial state,
even integrable systems always thermalize at long wavelengths, or, alternatively, can they exhibit genuinely non-equilibrium (critical) behavior?

In this manuscript, we consider the quench dynamics  of the prototypical Lipkin-Meshkov-Glick (LMG) model, an integrable model of spins with collective interactions.
We focus on the role of critical fluctuations and their universal properties. We show that, depending on the nature of the initial state (disordered or critical), distinct universal behaviors emerge in the dynamics. In particular, we show that fluctuations within the stationary state at late times can be described by an effective temperature which drastically depends on the initial state, and may vanish or even diverge for a quench from a critical state. We shall see that the latter divergence is a signature of a genuinely non-equilibrium critical behavior. 

\begin{figure}
\includegraphics[width=\linewidth]{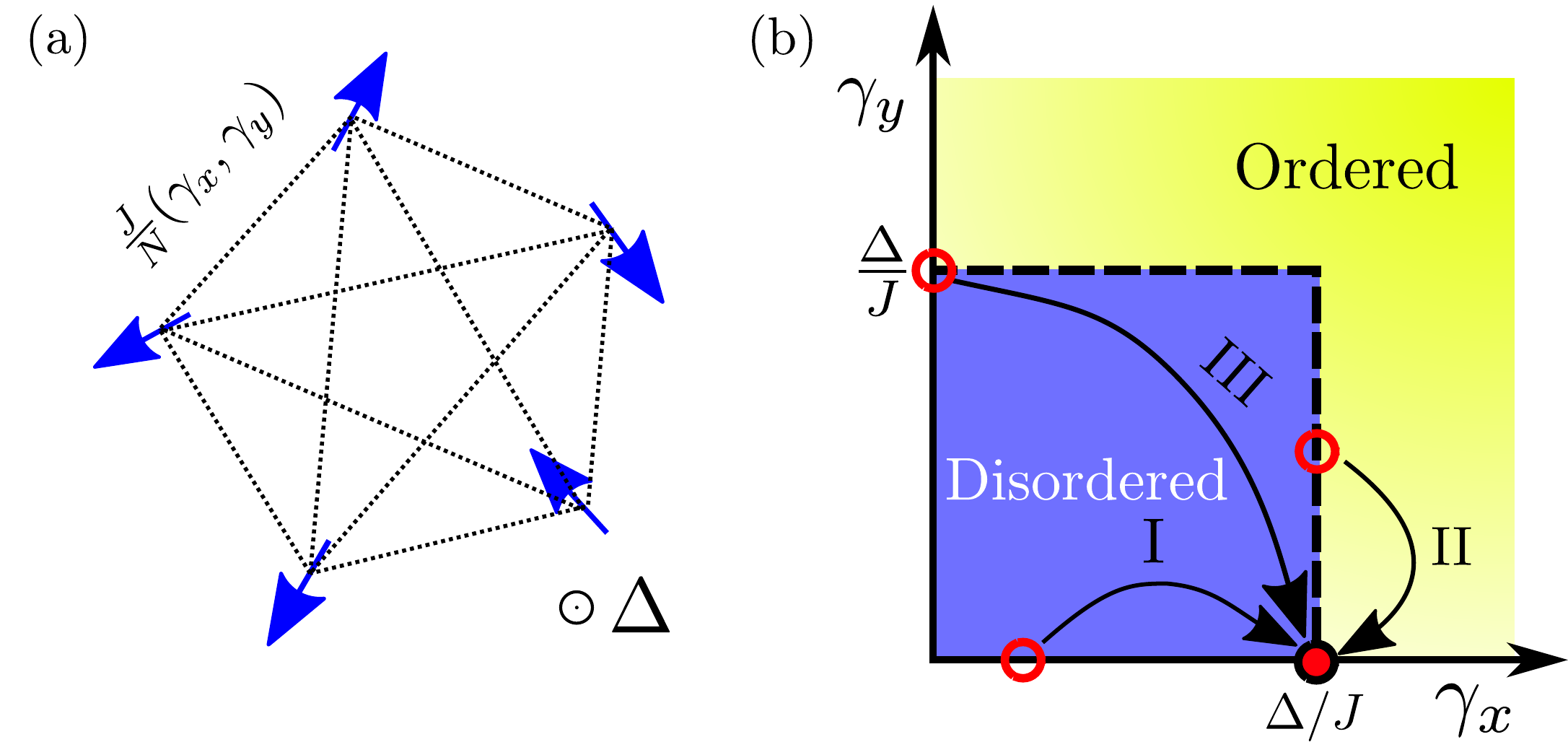}
\caption{(a) The LMG model [Eq.~\eqref{eq:model}] shown schematically. Spins interact with one another via an anisotropic XY Hamiltonian parametrized by $J(\gamma_x,\gamma_y)$ and are in a transverse magnetic field $\Delta$. 
(b) Ground-state phase diagram of the LMG model. The critical (dashed) line defines the disorder-to-order transition. 
The three different initial states, corresponding to Type-I, II and III quenches, are shown as open circles, and the quench Hamiltonian is denoted by a solid circle.}
\label{fig:schematic}
\end{figure}

We first emphasize what distinguishes our results from previous work. Conventionally, quench dynamics in the LMG model has been investigated with the spins initially in the ordered phase with the dynamics governed by mean-field equations ~\cite{Campbell16,Zunkovich16,Mori17, Zhang17,Homrighausen17,Zunkovich18}. In contrast, we consider quenches where the order parameter is zero and the dynamics of fluctuations cannot be captured by a mean-field treatment. Quench dynamics near critical points have also been explored extensively in the context of the Kibble-Zurek mechanism~\cite{Zurek1985,Kibble1976} as the system is ramped through a quantum critical point~\cite{Polkovnikov2005,Polkovnikov2008}. However, this mechanism requires the initial and final states to be close to each other and a critical point setting it apart from our work.

{\it Model.}---%
We consider a  prototypical model of infinite-range interactions known as the Lipkin-Meshkov-Glick (LMG) model~\cite{Lipkin1965,Lipkin1965a} which was originally introduced in nuclear physics, but has also been used to describe other physical systems such as Bose-Einstein condensates \cite{Cirac98}, small ferromagnetic particles \cite{Chudnovsky88} as well as  trapped ions~\cite{Britton2012,Bohnet2016,Zhang17}. The LMG Hamiltonian is given by
\begin{equation}
    H=-\frac{J}{N}\sum_{i<j}\gamma_{x}\sigma_{i}^{x}\sigma_{j}^{x}+\gamma_{y}\sigma_{i}^{y}\sigma_{j}^{y}-\Delta \sum_{i}\sigma_{i}^{z},
    \label{eq:model}
\end{equation}
with the XY-like anisotropic interaction characterized by $\gamma_{x,y}$ and the transverse field $\Delta$. The collective nature of the model allows us to write the Hamiltonian in terms of the total spin operators $S_a=\frac{1}{2}\sum_i \sigma^{a}_i$ with $a=x,y,z$. Note that the Hamiltonian commutes with $\bS^2=S_x^2+S_y^2+S_z^2$, making it block diagonal in a basis defined by the total spin $S$. In fact, this model is integrable \cite{Leboeuf90} and exactly solvable using Bethe ansatz \cite{Pan99,Morita06}. The model of interacting spins is  schematically shown in \cref{fig:schematic}(a).

The ground-state phase diagram of the LMG model [\cref{fig:schematic}(b)] is well known, exhibiting a transition from a disordered paramagnet to an ordered ferromagnet ~\cite{Botet_1983,Dusuel2004,Ribeiro2007,Ribeiro2008}. The in-plane magnetization serves as an order parameter where $\frac{1}{N}\langle S_{x}\rangle\neq 0$ and/or $\frac{1}{N}\langle S_{y}\rangle\neq 0$ in the ordered phase and $\frac{1}{N}\langle S_{x,y}\rangle=0$ in the disordered phase, in the thermodynamic limit $N\to \infty$. The two phases are separated by a continuous transition along $\Delta=J \,{\rm max}\{\gamma_x,\gamma_y\}$. Given the infinite-range interactions, the phase diagram can be obtained from a mean-field analysis. However, mean-field theory is insufficient where the order parameter is zero (outside the ordered phase) and particularly fails to capture the divergent fluctuations at the critical point. These fluctuations scale with $N$ as~\cite{Leyvraz2005,Vidal2006}
\begin{equation}
\frac{1}{N}\langle S_x^2\rangle\sim N^{1/3},\ \qquad \frac{1}{N}\langle S_y^2\rangle\sim N^{-1/3},  
\label{eq:GSfluctuations}
\end{equation}
along the critical line $\gamma_x>\gamma_y$; here, the prefactor $1/N$ factors out the trivial (square-root-volume) scaling away from criticality. Notice that the normalized spin fluctuations diverge only along the direction with the larger interaction strength. 
The exponent characterizing this divergence ($1/3$) is a distinct signature of the quantum phase transition~\cite{Leyvraz2005,Vidal2006}. 
For comparison, the fluctuations at a finite-temperature phase transition for this model diverge with a different critical exponent of 1/2 \cite{Paz19}. Therefore, the critical exponents distinguish between the quantum and thermal phase transitions.

{\it Quench dynamics.}---It is widely believed that the dynamics following a sudden quench leads to thermalization. The LMG model being integrable does not fully thermalize; nonetheless, we will see that a generic quench from the disordered phase gives rise to 
an effectively thermal behavior (including critical exponents) describing low-frequency modes. The question, however, remains if effective thermalization can be evaded at all, and, specifically, if a new, non-thermal scaling could emerge? Remarkably, the answer is in the affirmative. To show this, we study the dynamics for different types of quenches and initial states. To expose the critical behavior, the post-quench Hamiltonian is considered to be one at a critical point; without loss of generality, we take $\{\gamma_x=1,\gamma_y=0,\Delta=J\}$. We consider three different initial states, each corresponding to the ground state of the LMG Hamiltonian but at different parameters: (i)~\mbox{Type-I}: Initial state deep in the disordered phase, (ii)~\mbox{Type-II}: Critical initial state on the critical line $\Delta=\gamma_x J$, (iii)~\mbox{Type-III}: Critical initial state on the critical line $\Delta=\gamma_y J$; see \cref{fig:schematic}(b).
Types II and III are distinguished by their initial divergent fluctuations in $S_x$ and $S_y$, respectively.

It is instructive to first discuss a quench within  the disordered phase. In this regime, we can simplify the dynamics by using the Holstein-Primakoff approximation \cite{Das_2006}, $S_{z}=\frac{N}{2}-a^{\dagger}a$ and $S_x-iS_y \approx\sqrt{N} a$. Defining $a=\frac{1}{\sqrt{2}}(x+ip)$, we can write $S_x\approx\sqrt{N/2}\,x$ and $S_y\approx-\sqrt{N/2}\,p$. The LMG Hamiltonian \eqref{eq:model} can be then cast as a harmonic oscillator with the frequency and the mass defined as $\Omega^2\equiv 4\Delta^{2}\left(1-{J\gamma_x}/{\Delta} \right)\left(1-{J\gamma_y}/{\Delta} \right)$ and $2m\equiv{1}/(\Delta-\gamma_yJ)$, respectively; see the Supplemental Material (SM)~\cite{supp}. In this picture, the quench can be viewed as a sudden change of the mass and frequency of the oscillator, $\{m_0,\Omega_0\}\rightarrow \{m,\Omega\}$. It is straightforward to characterize the  fluctuations at long times in a quench to the vicinity of the critical point ($m\Omega\ll m_0\Omega_0$) \cite{supp}, 
\begin{equation}
    \frac{1}{N}\langle S_x^2(t)\rangle_{t\rightarrow\infty}=\frac{1}{2}\langle x^2(t)\rangle_{t\rightarrow\infty}\approx\frac{m_0\Omega_0}{8m^2\Omega^2}.
    \label{eq:HP-longtimeSx}
\end{equation}
This expression is reminiscent of a high-temperature harmonic oscillator ($T\gg \Omega$) where the equipartition theorem dictates $\frac{1}{2} m \Omega^2\langle x^2\rangle\approx \frac{1}{2} T$ which hints at the emergence of an effective temperature $T_{\rm eff}=m_0\Omega_0/4m$ \cite{Gambassi-Mitra15}, or, equivalently,  
\begin{equation}
T_{\rm eff}=\frac{\Delta}{2}\sqrt{\frac{\Delta_0-J\gamma_{x0}}{\Delta_0-J\gamma_{y0}}}.
\label{eq:Teff}
\end{equation}
Further insight can be obtained by examining the behavior of $T_{\rm eff}$ for the different quenches. For the Type-I quench, the initial state is disordered ($\Delta_0> J\gamma_{x0},J\gamma_{y0}$) giving rise to a finite effective temperature. For a Type-II quench, the initial state is critical ($\Delta_0=J\gamma_{x0}>J\gamma_{y0}$), resulting in a vanishing effective temperature. Most surprisingly, for the Type-III quench, the critical initial state  ($\Delta_0=J\gamma_{y0}>J\gamma_{x0}$) leads to a divergent effective temperature.
This simple analysis hints at qualitatively different behaviors in Type I, II and III, which we will identify with an effective thermal, quantum, and non-equilibrium critical behavior, respectively. To this end, we shall go beyond the Holstein-Primakoff approximation both numerically using exact diagonalization and analytically via the Keldysh field theory. 
 
First, we introduce universal scaling functions that capture the dynamics of the correlations and fluctuations,
\begin{subequations}\label{Eq. scaling}
\begin{align}
\frac{1}{N}\langle S_x^2(t)\rangle
&= N^\alpha f\left(\frac{t}{N^\zeta}\right)\label{eq:scalingform-1},  \\
C=\frac{1}{2N}\langle [S_x(t_2),S_x(t_1)]_+\rangle_{\rm st.}
&= N^\alpha  \,\tilde C\left(\frac{t_2-t_1}{N^\zeta}\right)\label{eq:scalingform-2}, \\
\chi=\frac{1}{2iN}\langle [S_x(t_2),S_x(t_1)]_-\rangle_{\rm st.} &= N^\zeta \,\tilde  \chi\left(\frac{t_2-t_1}{N^\zeta}\right),\label{eq:scalingform-3}
\end{align}
\end{subequations}
where $f, \tilde C, \tilde \chi$ are scaling functions. 
The two-time correlators $C$ and $\chi$ denote the correlation and response functions, respectively, which in equilibrium are related via the fluctuation-dissipation theorem \cite{Tauber14-book}. The subscript (st) indicates the long-time limit ($t_1,t_2\gg |t_1-t_2|$) when a stationary state is approached. We introduce two critical exponents: $\zeta$ defines a dynamical exponent for the scaling of the characteristic time scale of the dynamics with system size, and $\alpha$ characterizing the scaling of fluctuations. Remarkably, we shall see that the same exponents describe the entire dynamics both at short times as well as the stationary state at long times.

{\it Numerical results.}---%
Let us discuss the numerical results for the quench dynamics in the LMG model. The total-spin conservation allows us to simulate using exact diagonalization for large system sizes up to $N=9000$. For the numerical simulations, we set $J=1$, and restrict ourselves to the largest spin sector $S=N/2$. 
Beside the post-quench Hamiltonian with $\{\gamma_x=1,\gamma_y=0, \Delta=1\}$, the initial states correspond to the ground state of the Hamiltonian with the following parameters: (i)~Type-I: $\{\gamma_x=1,\gamma_y=0, \Delta=4\}$, (ii)~Type-II: $\{\gamma_x=1,\gamma_y=0.5, \Delta=1\}$, (iii)~Type-III: $\{\gamma_x=0,\gamma_y=1, \Delta=1\}$. 

\begin{figure}
\includegraphics[width=\linewidth]{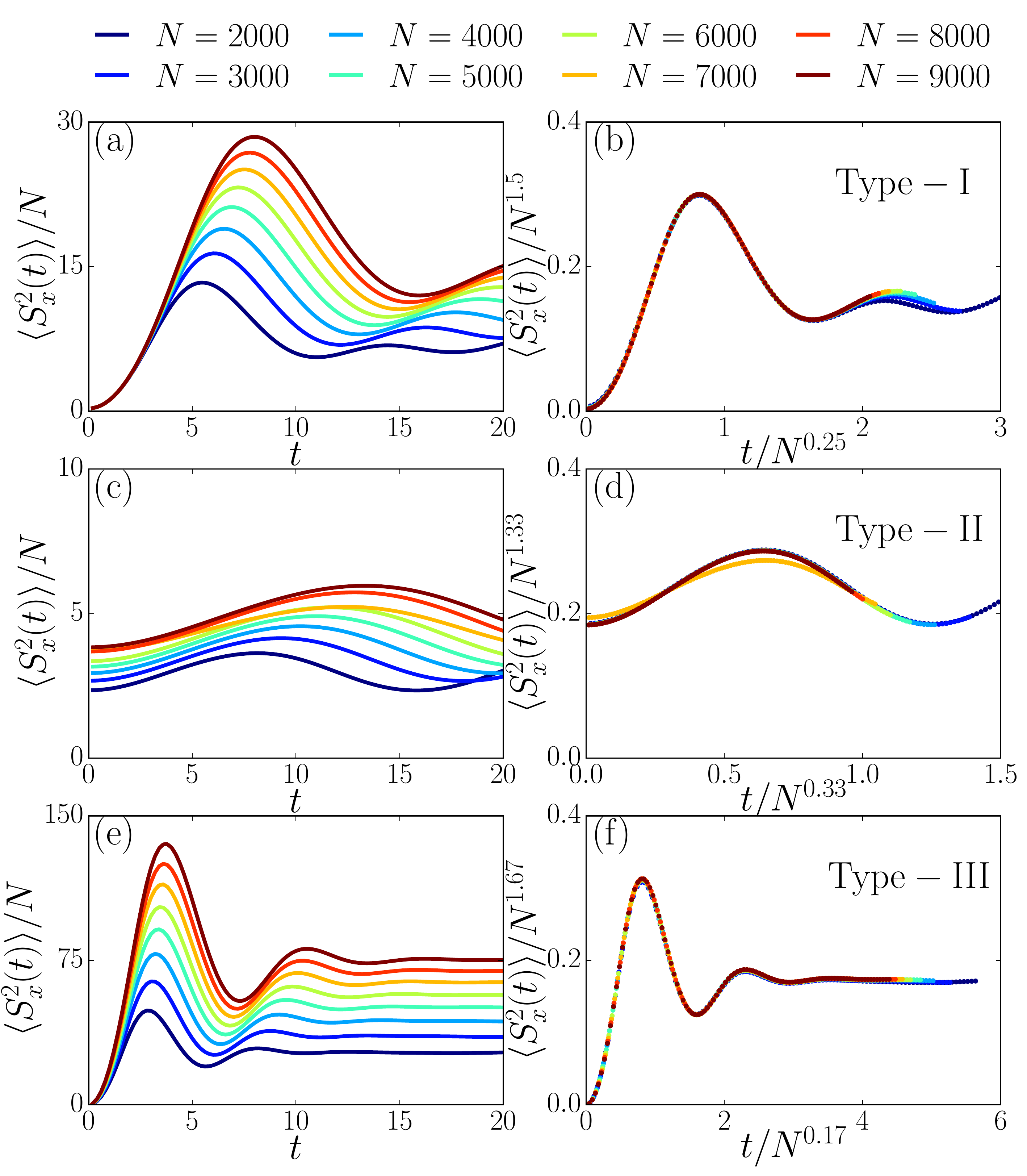}
\caption{Evolution of the fluctuations, $\frac{1}{N}\langle S_x^2\rangle$, for quench Types I, II and III (see Fig.~\ref{fig:schematic}) reported in the first, second and third rows, respectively. The right column shows the rescaled data which identifies the critical exponents $\alpha$ and $\zeta$ characterizing the overall scaling of fluctuations as well as the dynamics [see Eq.~\eqref{Eq. scaling}].}
\label{fig:fluctations}
\end{figure}

The fluctuations in the order parameter are shown in \cref{fig:fluctations} where each row corresponds to a given quench type. In the case of the Type-I quench [Figs.~\ref{fig:fluctations}(a) and (b)], the initial state is in the disordered phase with small (i.e., non-critical) fluctuations. Fluctuations grow initially ($t\lesssim1/J$) independent of the system size, but peak at longer times that increase with the system size. 
The scaling collapse of the different curves [Fig.~\ref{fig:fluctations}(b)] indicates that fluctuations diverge as $N^{0.5}$ and furthermore evolve with a characteristic time scale $\sim N^{0.25}$ before reaching the stationary state; hence, we identify the exponents $\zeta=0.25$ and $\alpha=0.5$. Indeed, the same exponents govern the two-time correlators in the stationary state consistent with the scaling in Eq.~\eqref{Eq. scaling}; see the SM \cite{supp}. Interestingly, these exponents are \textit{identical} to those governing a thermal critical point~\cite{Paz19}. This might be surprising because a true thermal phase transition only occurs at $\Delta/J<1$ \cite{Suzuki2013} in contrast with  $\Delta/J=1$ chosen above, hence the stationary state cannot be described as a Gibbs state \cite{Vidmar2016}.
However, the fact that the critical behavior is consistent with a thermal phase transition hints at an \textit{effective} thermalization at low frequencies. 

For the Type-II quench, the initial state is critical with divergent fluctuations in $S_x$. As shown in Fig. \ref{fig:fluctations}(c), fluctuations do not significantly grow over time. This observation indicates that the sudden quench has only slightly disturbed the system. Indeed, the scaling collapse shown in Fig. \ref{fig:fluctations}(d) reveals the exponents $\zeta=0.33$ and $\alpha=0.33$, consistent with a quantum critical behavior already present in the initial state. The emergence of quantum criticality in a quench from a critical state is also observed in the $O(N)$ model at large $N$~\cite{Chiocchetta17}.

\begin{figure}
\includegraphics[width=\linewidth]{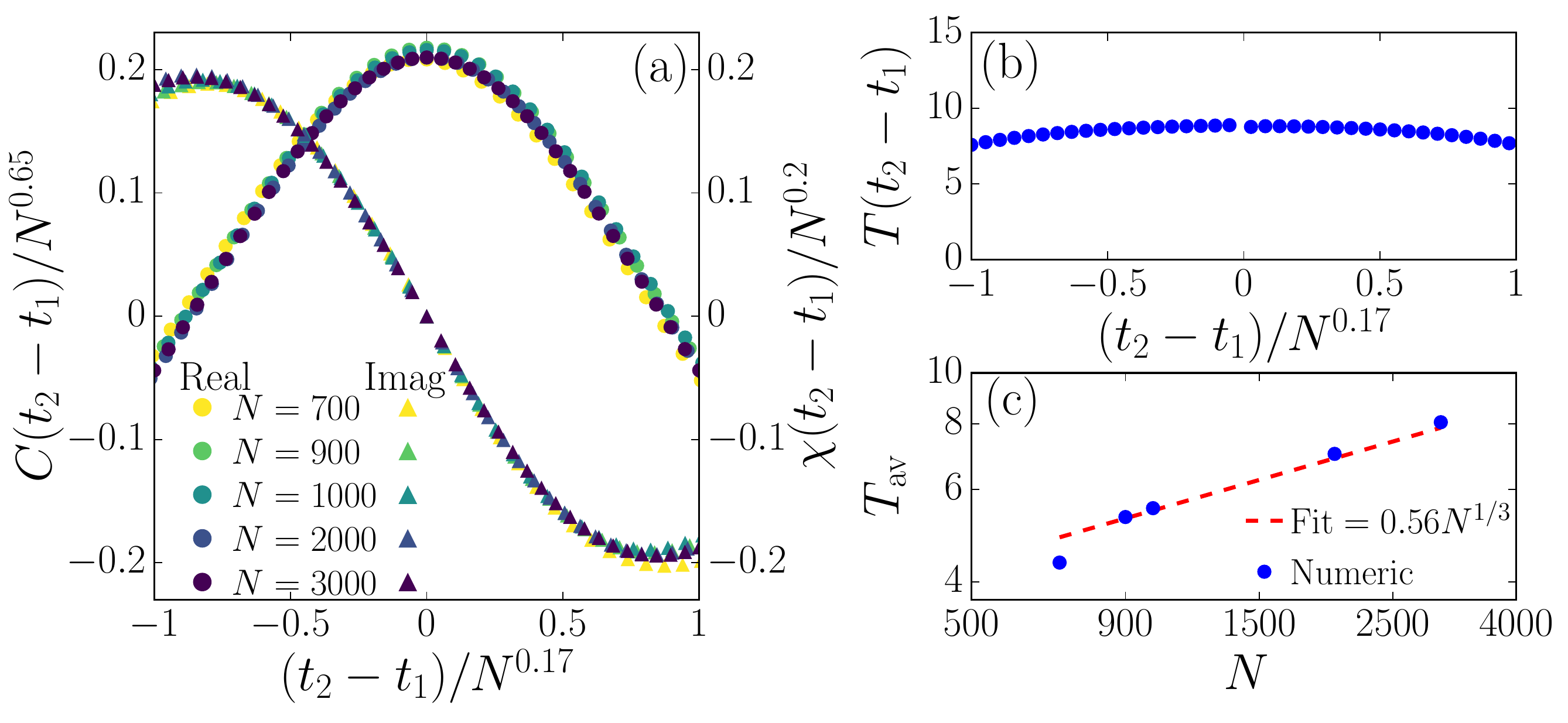}
\caption{(a) The correlation and response functions within the stationary state for the Type-III quench and for different system sizes. $t_2=20$ is chosen to ensure that the stationary state is reached.
A scaling collapse is consistent with the exponents $\zeta=1/6$ and $\alpha=2/3$.  (b) A time-dependent effective temperature, $T(t)$, extracted from the two-time correlators for $N=3000$ shown in (a). The data approximately fits a time-independent effective temperature. (c) The time-average effective temperature scales with the system size as $T_{\rm av}\sim N^{1/3}$.}
\label{fig:two-time}
\end{figure}

Most interestingly, the Type-III quench exhibits novel non-equilibrium behavior that is neither thermal nor quantum critical. As shown in \cref{fig:fluctations}(e), fluctuations grow faster than the other quenches. While the dynamics might seem similar to the Type-I quench, the exponents are markedly different: $(\zeta,\alpha)=(0.17,0.67)$. This indicates that fluctuations diverge with the system size even more strongly than those in Type-I or, equivalently, at the thermal critical point. Indeed, as we shall see shortly, the effective temperature in this case diverges with the system size. We also compute the two-time correlators for this quench as shown in \cref{fig:two-time}(a). Again, we find that the correlation and response functions obey the scaling forms in \cref{eq:scalingform-2,eq:scalingform-3} with approximately the same critical exponents.

Inspired by the fluctuation-dissipation theorem, we can also identify an effective temperature describing the stationary state. In equilibrium and at low frequencies (relative to temperature), this theorem relates fluctuations and dissipation as $C(\omega)=(2T/\omega) i\chi(\omega)$ \cite{Tauber14-book}. This equation can also be recast in the time domain,  $\chi(t)=\frac{1}{2T}\partial_t C(t)$ with $t$ the time difference. Away from equilibrium, this relation is commonly used to identify a time-dependent effective temperature \cite{Cugliandolo11}. Indeed, for the Type-III quench, the two-time correlators in \cref{fig:two-time}(a) satisfy the latter equation for an approximately constant $T(t)$; see \cref{fig:two-time}(b). In \cref{fig:two-time}(c), it is clear that the average temperature, $T_{\rm av}=\frac{1}{t}\int_0^t d\tilde{t}\ T(\tilde{t})$, scales with the system size as $T_{\rm av} \sim N^{1/3}$, and indeed diverges with the system size. 
This should be contrasted with the constant effective temperature for the Type-I quench \cite{supp}.We note that $T_{\rm av}$ is technically different from $T_{\rm eff}(\omega=0)$; the latter is evaluated from the IR limit of the ratio of correlation and response functions in the frequency domain which is difficult to access numerically. Nevertheless, we expect the scaling behavior to be the same. 

{\it Scaling analysis.}---%
The scaling behavior observed in the numerics hints towards an underlying scaling theory. In fact, this behavior and the scaling relations postulated in \cref{eq:scalingform-1,eq:scalingform-2,eq:scalingform-3} can be obtained from an effective low-frequency theory. At (or near) the critical point of the post-quench Hamiltonian, the relevant degree of freedom can be identified from the Holstein-Primakoff bosons as $x=(a+a^\dagger)/\sqrt{2}$, which also serves as the order parameter within the ordered phase. The non-equilibrium nature of the model necessitates the Keldysh formalism at the expense of doubling the fields $x\to x_{c/q}$ that define the Keldysh basis. The non-equilibrium partition function is given by~\cite{supp}
\begin{equation}
    Z=\int [dx_{c,q}]\,W\left(x_{c0}, K\dot x_{c0}\right) \, e^{iS_K}.
    \label{eq:Keldysh-partition-function}
\end{equation}
Here, $W(x_c,p_c)$ is the Wigner function describing the initial state (hence the variables with the subscript 0), and $S_K$ is the Keldysh action that is given by $S_K= -\int_0^\infty dt\,\left[K x_q \ddot x_c+ r x_q x_c +\frac{u}{N}(x_c^2+x_q^2)x_cx_q\right]$. The coefficients in the action are given in terms of the microscopic parameters as 
$K^{-1} \equiv 2(\Delta-J\gamma_y)$, $r\equiv 2( \Delta-J\gamma_x)$ and $u\equiv J \gamma_x$; see the SM \cite{supp}. Notice that $r= 0$ defines the critical point after the quench. The Wigner function describing the initial state can be written as $W(x,p)={\cal W}(x^2N^{-\alpha_0},p^2N^{\alpha_0})$ where $\alpha_0=0, 1/3,-1/3$ for Type I, II and III, respectively. This correctly produces $\langle x^2\rangle \sim N^{\alpha_0}$ (similarly for $p$) in the initial state; see \cref{eq:GSfluctuations}. The precise form of the Wigner function as well as the coefficients in the action are not required for the scaling analysis; for convenience, we set $K=1$.

\begin{table}[t]
\centering
\begin{tabular*}{0.8\linewidth}{c@{\extracolsep{\fill}}||ccccc}
        \hspace{0.35in}  & Type I & Type II & Type III  & TCP & QCP  \\ \hline\hline 
$\zeta$ & $\frac{1}{4}$ & $\frac{1}{3}$ & $\frac{1}{6}$ & $\frac{1}{4}$ & $\frac{1}{3}$  \\[2pt] 
$\alpha$  & $\frac{1}{2}$ & $\frac{1}{3}$ & $\frac{2}{3}$ & $\frac{1}{2}$ &  $\frac{1}{3}$   \\[2pt] 

\hline
$T_{\rm eff}^{\rm IR}$ & $\rm finite$ & $0$ & $\sim N^{1/3}$ & $T_c$ & 0 \\[2pt] 
 \hline
\end{tabular*}
\caption{The critical exponents $\zeta$ and $\alpha$ characterizing the scaling of dynamics and fluctuations, respectively. Distinct critical behaviors emerge in quench Types I, II and III. Types I and II give rise to identical exponents as thermal and quantum critical points (TCP/QCP), respectively. Genuinely non-equilibrium exponents emerge in Type III. The IR limit ($\omega\rightarrow 0$) of the effective temperature is finite, zero, and divergent for the three types of quench, respectively.} 
\label{tab:exponents}
\end{table}

We shall focus on Type I and III where the system is significantly disturbed upon quench. 
At the critical point, $r=0$, we expect (confirmed by numerics) that the fluctuations diverge, that is, $\frac{1}{N}\langle S_x^2\rangle\sim\langle x_c^2 \rangle \sim N^{\alpha}$ for an exponent $\alpha(>0)$ to be determined; we then identify the scaling dimension $[x_c]=\frac{\alpha}{2}$. This divergence can be scaled away by defining $X_c=x_c N^{-\alpha/2}$. Anticipating (again confirmed by numerics) a critical slowdown, we also introduce a rescaled time $T =t N^{-\zeta}$ with a dynamical exponent $\zeta$. By appropriately choosing these exponents, the action together with the Wigner function can be made scale invariant, i.e., independent of $N$. Writing the Wigner function in terms of the rescaled variables, we have $W(x_{c0},  \dot x_{c0})={\cal W} \left(X_{c0}^2 N^{\alpha-\alpha_0},  {X_{c0}'}^{\!\!2} N^{\alpha-2\zeta+\alpha_0}\right)$ with $X'=dX/dT$. 
Now, the function $\cal W$ is only significant when its arguments are of order 1. Therefore, fluctuations of $X_{c0}$ are greatly suppressed for Type-I and III quenches ($\alpha_0\le 0$), thus imposing $X_{c0}\approx 0$. On the other hand, fluctuations of $X_{c0}'$ can be made scale invariant by setting $\alpha-2\zeta+\alpha_0=0$. Similar analysis can be made for the terms in the action. The kinetic term ($\sim\int_t x_q \ddot x_c$) can be made scale-invariant by introducing the new rescaled variable  $X_q=N^{-\alpha/2+\zeta} x_q$, hence the scaling dimension $[x_q]=-\frac{\alpha}{2}+\zeta$. Next, we turn to the interaction term, $(-1/N)\int_t u_c x_q x_c^3+u_q x_q^3 x_c$, where ``classical'' ($\sim u_c$) and ``quantum'' ($\sim u_q$) vertices are distinguished although $u_c=u_q$ at the microscopic level. Let us first cast the classical vertex in terms of the rescaled variables, upon which $u_c/N \to u_c N^{2\zeta+\alpha-1}$. Making the latter scale invariant yields $2\zeta+\alpha-1=0$. Combining the above relations, the critical exponents are determined as $\zeta=\frac{1+\alpha_0}{4}$ and $\alpha=\frac{1-\alpha_0}{2}$. 
One can then confirm that, upon the scaling transformation, the quantum vertex is suppressed by a power of $N$ and thus can be neglected; in the language of renormalization group theory, the quantum vertex is irrelevant. This itself is a consequence of classical and quantum fields having different scaling dimensions ($[x_c]=\frac{1-\alpha_0}{4}$ vs $[x_q]=\frac{\alpha_0}{2}$). Our scaling analysis yields the exponents reported in  Table~\ref{tab:exponents} consistent with numerics, and also reproduces the behavior of the effective temperature (see the SM \cite{supp}). We remark that, in analogy with boundary critical phenomena \cite{Diehl81,Diehl86}, our analysis has relied on scaling both ``boundary'' and ``bulk'' terms. 

While we have focused on finite-size scaling, we can just as well identify the scaling behavior in the thermodynamic limit ($N\to \infty$) away from the critical point ($r>0$). Interestingly, we find that the fluctuations at late times diverge as $1/r$, $1/\sqrt{r}$ and $1/r^2$ for Type I, II and III, respectively; see the SM \cite{supp}. While the first two scalings can be identified with the characteristic quantum and thermal behavior, the distinct scaling for Type III is indicative of genuinely non-equilibrium critical behavior.

The non-equilibrium dynamics reported in this work is accessible in a variety of experimental platforms, particularly in the context of trapped ions \cite{Blatt2012,Britton2012,Zhang17}. These systems are described by spin models with long- or even infinite-range interactions. 
A challenge is to prepare an initial critical state for Type-II and III quench dynamics. However, based on a quantum approximate optimization protocol \cite{Farhi14}, variational quantum algorithms have been recently proposed \cite{Ho19} and implemented \cite{Zhu-Monroe19} to efficiently prepare quantum critical states.

{\it Conclusion and outlook}---%
In this work, we  have studied quench dynamics in the context of infinite-range spin models. We have identified a rich dynamical universal behavior depending on the critical nature of the initial state. Specifically, we have identified a scenario (Type III) where a genuinely non-equilibrium critical behavior emerges both in the short-time dynamics and long-time stationary state. Our findings complement the distinct signatures of correlated initial conditions on the dynamics of classical systems \cite{Odor04}. 
An important future direction is to identify efficient routes to preparing critical initial states in such models. An intriguing alternative is to first perform a Type-I quench, whose initial state can be easily prepared, that leads to a (thermal-like) critical state at late times. Upon a second quench of Type III, the system then approaches another critical, yet non-equilibrium state at late times. Another particularly relevant direction is to extend our results to long-range spin models, $V(r)\sim 1/r^p$, specifically with $p<1$ where long-lived prethermal states emerge \cite{Mori19}. More generally, extending our results to other integrable models or those exhibiting long-lived prethermalization is worthwhile. 
Finally, it would be interesting to identify any \textit{ageing} behavior \cite{Gambassi05review} in such models. 

\begin{acknowledgments}
{\it Acknowledgments.}---%
We thank Marcos Rigol and Daniel Paz for useful discussions. P.T. acknowledges support from the NIST NRC Research Postdoctoral Associateship Award. M.M. acknowledges support
from NSF under Grant No. DMR-1912799 and start-up funding from Michigan State University. This work is supported in  part  by  the  U.S.  Department  of  Energy (DOE),  Office  of  Science,  Office  of  Advanced  Scientific Computing Research (ASCR) Quantum Computing Application Teams program, under fieldwork proposal number ERKJ347. The authors acknowledge the University of Maryland supercomputing resources (http://hpcc.umd.edu) made available for conducting the research reported in this Letter. This research was supported in part by the National Science Foundation under Grant No. NSF PHY-1748958.
\end{acknowledgments}

\bibliographystyle{apsrev4-1}
\bibliography{LMGrefs-combined}



\section*{Supplementary material}

# Supplemental Material

In this Supplemental Material, we outline the details omitted in the main text. In Sec. S.I, we discuss the Holstein-Primakoff approximation to discuss the spin dynamics in terms of bosons. In Sec. S.II, we review the Keldysh field theory that describes the dynamics of the LMG model. In Sec. S.III, we derive the scaling relations and critical exponents from a systematic scaling theory based on the Keldysh action. Finally, in Sec. S.IV, we derive the scaling behavior of the effective temperature using the fluctuation-dissipation relation for different types of quench.

## S.I. HOLSTEIN-PRIMAKOFF APPROXIMATION

In this section, we derive the dynamics of the total spin in terms of bosonic operators using the Holstein-Primakoff transformation. We will expand the Hamiltonian to the quadratic order in the bosonic operators and will also include the first $1/N$ correction that characterizes the nonlinear interactions. The bosonic model at the quadratic order already captures the main features provided that the initial and final states are in the disordered phase. The nonlinear terms are included for our discussion of criticality and finite-size scaling in Sec. S.II. We start with the Hamiltonian in Eq. (1) of the main text ($S_a = \frac{1}{2}\sum_i \sigma_i^a$),

$$H = -\frac{2J}{N}\left[\gamma_x S_x^2 + \gamma_y S_y^2\right] - 2\Delta S_z. \tag{S1}$$

In the Holstein-Primakoff representation, spins can be written in terms of bosonic operators as

$$S_z = \frac{N}{2} - a^\dagger a, \tag{S2a}$$

$$S_- = S_x - iS_y = \sqrt{N - a^\dagger a}\, a = \sqrt{N}(1 - \frac{a^\dagger a}{2N})a + \cdots, \tag{S2b}$$

$$S_+ = S_x + iS_y = a^\dagger \sqrt{N - a^\dagger a} = \sqrt{N}\, a^\dagger(1 - \frac{a^\dagger a}{2N}) + \cdots. \tag{S2c}$$

The interaction term in the Hamiltonian (S1) then reads in terms of Holstein-Primakoff bosons as

$$\frac{1}{N}\left[\gamma_x S_x^2 + \gamma_y S_y^2\right] = \gamma_x \left\{\frac{1}{4}(a+a^\dagger)^2 - \frac{1}{8N}\left[(a^\dagger + a)a^\dagger(a^\dagger + a)a + a^\dagger(a^\dagger + a)a(a^\dagger + a)\right] + \cdots\right\}$$
$$+ \gamma_y \left\{-\frac{1}{4}(a - a^\dagger)^2 + \frac{1}{8N}\left[(a^\dagger - a)a^\dagger(a^\dagger - a)a + a^\dagger(a^\dagger - a)a(a^\dagger - a)\right] + \cdots\right\}.$$

This expression is not normal- (nor, Weyl-) ordered. But, any such ordering requires commuting $a$ and $a^\dagger$ which will result in either a constant term or a quadratic term in $a$ and $a^\dagger$ that is nevertheless suppressed by $1/N$. We will therefore ignore the ordering altogether, and write

$$\frac{1}{N}\left[\gamma_x S_x^2 + \gamma_y S_y^2\right] = \frac{1}{4}\left[\gamma_x(a^\dagger + a)^2 - \gamma_y(a^\dagger - a)^2\right] + \frac{1}{N}(\text{quadratic in } a \text{ and } a^\dagger)$$
$$- \frac{1}{4N}\left[\gamma_x a^\dagger a(a^\dagger + a)^2 - \gamma_y a^\dagger a(a^\dagger - a)^2\right].$$

We also define the (position and momentum) variables

$$a = \frac{1}{\sqrt{2}}(x + ip), \quad a^\dagger = \frac{1}{\sqrt{2}}(x - ip). \tag{S3}$$

Here and in the subsequent analysis, we set $\hbar = 1$. In particular, we have $a^\dagger a + 1/2 = \frac{1}{2}(x^2 + p^2)$ as well as $(a + a^\dagger)/\sqrt{2} = x$ and $(a^\dagger - a)/\sqrt{2} = -ip$. The entire Hamiltonian then reads (ignoring an unimportant additive constant and the quadratic terms of the order $1/N$)

$$H = (\Delta - J\gamma_x)x^2 + (\Delta - J\gamma_y)p^2 + \frac{J}{2N}(x^2 + p^2)\left[\gamma_x x^2 + \gamma_y p^2\right]. \tag{S4}$$

At the quadratic level, the Hamiltonian can be written as a harmonic oscillator, $H = \frac{1}{2m}p^2 + \frac{1}{2}m\Omega^2 x^2$ with $\Omega^2 = 4(\Delta - J\gamma_x)(\Delta - J\gamma_y)$ and $m = \frac{1}{2(\Delta - J\gamma_y)}$.

## A. Quench dynamics at the quadratic level

Consider a quench from an initial Hamiltonian $H_0$ defined by the parameters $\{\gamma_{x0}, \gamma_{y0}, \Delta_0\}$ to the final Hamiltonian in Eq. (S1). This corresponds to a quench of the harmonic-oscillator Hamiltonian in the form of $\{\Omega_0, m_0\} \to \{\Omega, m\}$ with the values of the mass and the frequency given in terms of the corresponding microscopic parameters. In the Heisenberg representation, the operators $x$ and $p$ evolve as

$$x(t) = x(0) \cos \Omega t + \frac{p(0)}{m\Omega} \sin \Omega t, \tag{S5a}$$

$$p(t) = -m\Omega x(0) \sin \Omega t + p(0) \cos \Omega t. \tag{S5b}$$

The fluctuations in the initial state are determined from the equal distribution of the ground-state energy, $\frac{1}{2}\Omega_0$, between kinetic and potential energies; i.e., $\frac{1}{2m_0}\langle p^2(0)\rangle = \frac{1}{2}m_0\Omega_0^2\langle x^2(0)\rangle = \frac{1}{4}\Omega_0$. Utilizing these equations, we find

$$\langle x^2(t)\rangle = \frac{m_0\Omega_0}{4m^2\Omega^2}\left[\left(1 + \frac{m^2\Omega^2}{m_0^2\Omega_0^2}\right) + \left(\frac{m^2\Omega^2}{m_0^2\Omega_0^2} - 1\right)\cos 2\Omega t\right]. \tag{S6}$$

Using the Holstein-Primakoff transformation, it is then straightforward to obtain the steady-state value of spin fluctuations,

$$\lim_{t\to\infty}\langle S_x^2(t)\rangle = \frac{N}{2}\lim_{t\to\infty}\langle x^2(t\to\infty)\rangle = \frac{Nm_0\Omega_0}{8m^2\Omega^2}\left(1 + \frac{m^2\Omega^2}{m_0^2\Omega_0^2}\right), \tag{S7}$$

where we have dropped the highly oscillatory terms at long times.

The two-time correlators can also be obtained from Eq. (S5) as

$$\langle[x(t_2), x(t_1)]_+\rangle = \frac{m_0\Omega_0}{2m^2\Omega^2}\left[\left(1 + \frac{m^2\Omega^2}{m_0^2\Omega_0^2}\right)\cos[\Omega(t_2 - t_1)] + \left(\frac{m^2\Omega^2}{m_0^2\Omega_0^2} - 1\right)\cos[\Omega(t_1 + t_2)]\right], \tag{S8a}$$

$$\langle[x(t_2), x(t_1)]_-\rangle = -\frac{2i}{m\Omega}\sin[\Omega(t_2 - t_1)]. \tag{S8b}$$

We can then identify the correlation and response functions of the spin variables, $\frac{1}{N}\langle S_x(t_2)S_x(t_1)\rangle = C(t_2, t_1) + i\chi(t_2, t_1)$, by identifying $C(t_2, t_1) = \frac{1}{4}\langle[x(t_2), x(t_1)]_+\rangle$ and $\chi(t_2, t_1) = \frac{-i}{4}\langle[x(t_2), x(t_1)]_-\rangle$ via the Holstein-Primakoff transformation. In the limit of long times $t_1, t_2 \gg |t_2 - t_1|$, we find

$$\lim_{t_1,t_2\gg|t_2-t_1|} C(t_2, t_1) = \frac{m_0\Omega_0}{8m^2\Omega^2}\left(1 + \frac{m^2\Omega^2}{m_0^2\Omega_0^2}\right)\cos[\Omega(t_2 - t_1)], \tag{S9a}$$

$$\lim_{t_1,t_2\gg|t_2-t_1|} \chi(t_2, t_1) = -\frac{1}{2m\Omega}\sin[\Omega(t_2 - t_1)]. \tag{S9b}$$

Again, we have dropped the highly oscillatory terms at long times. Note that the two-time correlators only depend on the time-difference, $t_2 - t_1$, indicating that the time-translation symmetry, broken explicitly by the quench, is restored at long times.

## B. Effective temperature

In thermal equilibrium, the relation between correlation and response functions is given by the fluctuation-dissipation theorem [S1] as (briefly restoring $\hbar$)

$$C(\omega) = \hbar \coth\left(\frac{\hbar\omega}{2T}\right) i\chi(\omega). \tag{S10}$$

At low frequencies compared to the temperature ($\hbar|\omega| \ll T$), we recover the *classical* form of the fluctuation-dissipation relation, $C(\omega) = (2T/\omega)i\chi(\omega)$, where $\hbar$ goes away. Using the latter form of the fluctuation-dissipation relation (appropriate at low frequencies compared to temperature), we can extract the effective temperature from the correlation and response functions in Eq. (S9); we obtain

$$T_{\text{eff}} \approx \frac{m_0\Omega_0}{4m}\left(1 + \frac{m^2\Omega^2}{m_0^2\Omega_0^2}\right) \approx \frac{m_0\Omega_0}{4m}, \tag{S11}$$

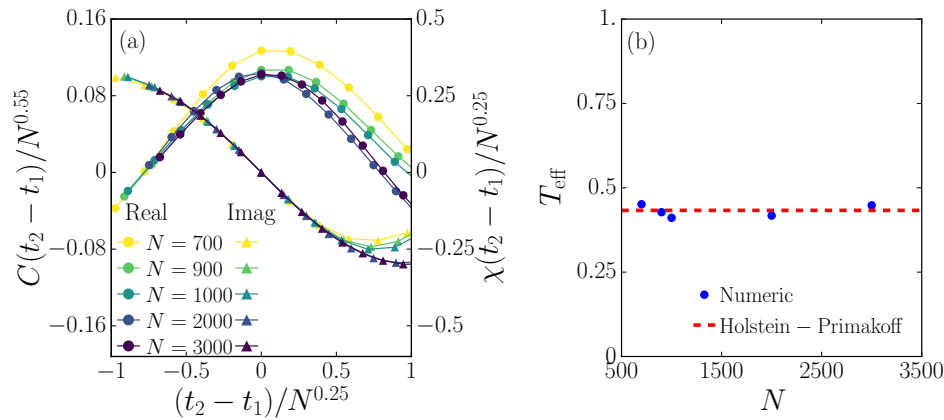

FIG. S1. (a) The correlation and response functions for the Type-I quench for different system sizes. A scaling collapse is achieved with the dynamical exponent $\zeta = 0.25$ and the exponent $\alpha = 0.55$ characterizing fluctuations. (b) Effective temperature $T_{\text{eff}}$ extracted from the two-time correlators shown in (a) via the fluctuation-dissipation relation. The data is consistent with a constant effective temperature independent of system size and agrees well with the effective temperature extracted from the Holstein-Primakoff approximation, $T_{\text{eff}} = \frac{m_0 \Omega_0}{4m} = \frac{\Delta}{2}\sqrt{\frac{(\Delta_0 - J)}{\Delta_0}} \approx 0.43$ for $\gamma_x = 1$, $\gamma_y = 0$, $\Delta_0 = 4$, $\Delta = 1$ and $J = 1$; see the dashed line.

where the last equality follows from our low-frequency assumption requiring $\Omega \ll T_{\text{eff}}$, which in turn implies $m\Omega \ll m_0 \Omega_0$. Considering a near-critical final state such that the Holstein-Primakoff approximation is still applicable, this condition is trivially satisfied in the Type-I quench when the initial state is deep in the disordered phase. Furthermore, assuming a nearly (but not exactly) critical initial state, this condition is also satisfied for Type III, but indicates a divergent effective temperature as the critical point is approached. For Type II, the above condition is never satisfied, indicating that the classical limit of the fluctuation-dissipation relation is not appropriate. This is consistent with our observation that the quantum critical behavior persists in the Type-II quench.

We also remark that the effective temperature identified via the fluctuation-dissipation relation is consistent with our analysis in the main text where the effective temperature is obtained by only inspecting the fluctuations. This is because the equipartition theorem is restored at long times, resulting in

$$\lim_{t \to \infty} \langle x^2(t) \rangle = \frac{T_{\text{eff}}}{m\Omega^2}. \tag{S12}$$

Comparing the above equation with Eq. (S6) while dropping the highly oscillatory terms, we recover the same effective temperature (S11).

The above analysis based on the Holstein-Primakoff approximation is appropriate away from criticality. Here, we show that, for the Type-I quench, the same effective temperature also relates the correlation and response functions at the critical point. In Fig. S1(a), we report the correlation and the response functions as a function of the time difference at long times and for different system sizes. The rescaled two-time correlators as a function of the rescaled time clearly converge for larger system sizes. In Fig. S1(b), the effective temperature, extracted from the fluctuation-dissipation relation, is shown to be constant (not scaling with $N$) and consistent with the prediction of the Holstein-Primakoff reported above (see also the main text). A similar analysis for the Type-III quench is presented in Fig. 3 of the main text, in which case it is shown that the effective temperature diverges with the system size. The scaling behavior of the effective temperature in the latter quench (Type III) is discussed in Sec. S.IV.

## S.II. SCHWINGER-KELDYSH PATH INTEGRAL

In this section, we cast the dynamics in terms of the Schwinger-Keldysh path integral. To this end, we resort to the coherent-state path integral and follow the conventions in Ref. [S2]. The non-equilibrium partition function reads as

$$Z = \int D\psi D\varphi \, W(\psi_0) e^{iS_K}, \tag{S13}$$

where $W$ represents the Wigner function of the initial state and the Keldysh action $S_K$ is given by

$$S_K = 2i \int_0^\infty dt \left( \bar{\varphi} \partial_t \psi - \varphi \partial_t \bar{\psi} \right) - \int_0^\infty dt \left[ \mathcal{H}(\psi + \varphi) - \mathcal{H}(\psi - \varphi) \right]. \tag{S14}$$

The symbol $\mathcal{H}$ represents the Weyl-ordered Hamiltonian in the coherent-state representation. However, as remarked in Sec. S.I, we can ignore this ordering. We identify $\psi + \varphi = a_+$ and $\psi - \varphi = a_-$. The usual Keldysh representation is defined as $a_{c/q} = (a_+ \pm a_-)/\sqrt{2}$ in terms of the fields on the forward and backward branches. We shall express these fields in terms of phase-space coordinates via $a_{c/q} = (x_{c/q} + ip_{c/q})/\sqrt{2}$. In short, we define

$$\psi = \frac{1}{2}(x_c + ip_c), \quad \varphi = \frac{1}{2}(x_q + ip_q). \tag{S15}$$

Next, we will express the action and the initial state in terms of the phase-space coordinates. The first term in the action Eq. (S14) reads

$$2i(\bar{\varphi}\partial_t \psi - \varphi \partial_t \bar{\psi}) = p_q \partial_t x_c - x_q \partial_t p_c.$$

The above transformation also allows us to write the Hamiltonian term in the action. To this end, we take the functional form $\mathcal{H}(x,p)$ of the Hamiltonian in Eq. (S4) (again we shall not worry about the Weyl ordering) and construct

$$\mathcal{H}\left(\frac{x_c + x_q}{\sqrt{2}}, \frac{p_c + p_q}{\sqrt{2}}\right) - \mathcal{H}\left(\frac{x_c - x_q}{\sqrt{2}}, \frac{p_c - p_q}{\sqrt{2}}\right) = 2(\Delta - J\gamma_x)x_c x_q + 2(\Delta - J\gamma_y)p_c p_q$$
$$+ \frac{J}{2N}\left[(x_c^2 + x_q^2 + p_c^2 + p_q^2)(\gamma_x x_c x_q + \gamma_y p_c p_q) + (x_c x_q + p_c p_q)\left(\gamma_x(x_c^2 + x_q^2) + \gamma_y(p_c^2 + p_q^2)\right)\right].$$

Here, the second line represents the nonlinear interactions. Assuming that $\Delta - J\gamma_y > 0$ in the post-quench state without loss of generality, the first line in this equation indicates that $p$ is effectively massive. We nevertheless let $\Delta - J\gamma_x$ to be zero or to be tuned near criticality. We can then simply drop the nonlinear $p$-dependent terms in the action. This will lead to a rather simple interaction term in the action

$$S_{\text{int}} \approx -\frac{J\gamma_x}{N} \int_0^\infty dt\, (x_c^2 + x_q^2)x_c x_q + \text{irrelevant terms}. \tag{S16}$$

We can then write the total action as

$$S_K = \int_0^\infty dt \Big[ p_q \partial_t x_c - x_q \partial_t p_c - 2(\Delta - J\gamma_x)x_c x_q - 2(\Delta - J\gamma_y)p_c p_q - \frac{J\gamma_x}{N}(x_c^2 + x_q^2)x_c x_q \Big]. \tag{S17}$$

The $p$-dependent terms at the quadratic level may be integrated out too. To this end, we first make an integration by parts

$$\int_0^\infty dt\, x_q \partial_t p_c = -x_{q0} p_{c0} - \int_0^\infty dt\, p_c\, \partial_t x_q.$$

Since the action is at most quadratic in $p$, we can use the standard saddle-point equation

$$\frac{\delta S_K}{\delta p_{c/q}(t)} = 0 \quad \longrightarrow \quad p_{c/q}(t) = K \partial_t x_{c/q},$$

where we have defined $K^{-1} \equiv 2(\Delta - J\gamma_y)$. This will result in the action

$$S_K = -x_{q0}p_{c0} + \int_0^\infty dt\, \Big[K\dot{x}_q \dot{x}_c - r x_q x_c - \frac{u}{N}(x_c^2 + x_q^2)x_c x_q\Big], \tag{S18}$$

where we have defined $r \equiv 2(\Delta - J\gamma_x)$ and $u \equiv J\gamma_x$. The full path integral can be now written as

$$Z = \int Dx_{c/q}(t) dp_{c0} W(x_{c0}, p_{c0}) e^{iS_K}, \tag{S19}$$

with the action given in Eq. (S18). Notice that the boundary term still depends on $p_{c0}$. A further integration by parts turns $\int_t \dot{x}_q \dot{x}_c = -x_{q0}\dot{x}_{c0} - \int_t x_q \ddot{x}_c$. The boundary term in this equation can be combined with that in Eq. (S18), which upon integration over $x_{q0}$ leads to a delta function that sets $p_{c0} = K\dot{x}_{c0}$. This will result in Eq. (6) in the main text together with the action reported thereafter.

Finally, we consider the Wigner function in the functional path integral. A familiar example of the Wigner function is that of harmonic oscillator applicable to an initial state deep in the disordered phase. In this limit, the initial Hamiltonian is $H_0 = a^\dagger a = \frac{1}{2}(x^2 + p^2)$; here, we have conveniently set the mass and frequency to 1. The Wigner function for this state is simply a Gaussian function given by $W(x,p) = 2e^{-x^2-p^2}$ in units that $\hbar = 1$ [S3]. The Wigner function in the path integral should be written in terms of $x_{c0} = x(0)/\sqrt{2}$ and $p_{c0} = p(0)/\sqrt{2}$ as

$$W(x_{c0}, p_{c0}) = e^{-x_{c0}^2/2 - p_{c0}^2/2}. \tag{S20}$$

In the partition function, we should replace $p_{c0}$ via $p_{c0} = K\dot{x}_{c0}$. In the subsequent analysis, however, we will not specify the exact form of the initial state, and consider a general scaling form describing different initial states. Following the convention in the main text, we write the partition function (setting $K=1$ for convenience)

$$Z = \int Dx_{c/q} \mathcal{W}\left(\tau_0 x_{c0}^2 N^{-\alpha_0}, \dot{x}_{c0}^2 N^{\alpha_0}\right) e^{-\int_0^\infty dt \left[x_q \ddot{x}_c + r x_q x_c + \frac{u_c}{N} x_c^3 x_q + \frac{u_q}{N} x_c x_q^3\right]}, \tag{S21}$$

where we have written the interaction term as the sum of classical ($\sim u_c$) and quantum ($\sim u_q$) vertices, although $u_c = u_q = u$. Furthermore, we have introduced the constant $\tau_0$ in the initial state although we can set $\tau_0 = 1$. We also recall that $\alpha_0 = 0, 1/3, -1/3$ for Type I, II and III, respectively.

## S.III. SCALING ANALYSIS & CRITICAL EXPONENTS

Equipped with the non-equilibrium partition function and the corresponding Schwinger-Keldysh action derived in Sec. S.II, we can now inspect the critical behavior of the dynamics. In the subsequent analysis, we shall focus on Type I and III where $\alpha_0 \leq 0$. We can identify a scale-invariant fixed point (for a general $\alpha_0 \leq 0$) by scaling various fields and parameters in the partition function (S21) as

$$t \to t/\lambda, \quad r \to \lambda^2 r, \quad x_c \to x_c/\lambda^{\frac{1-\alpha_0}{1+\alpha_0}}, \quad x_q \to x_q/\lambda^{\frac{2\alpha_0}{1+\alpha_0}}, \quad N \to N/\lambda^{\frac{4}{1+\alpha_0}}. \tag{S22}$$

Under this rescaling, all the terms remain scale invariant except those involving $\tau_0$ and $u_q$ which will flow to (as $\lambda \to \infty$)

$$\tau_0 \to \infty, \qquad u_q \to 0. \tag{S23}$$

The flow of $\tau_0$ imposes the "Dirichlet boundary condition" on $x$ at $t=0$. To illustrate this point, let's first consider Type I in which case the dependence on the initial position is given by $e^{-\tau_0 x_{c0}^2/2}$; the scaling behavior at the fixed point ($\tau_0 \to \infty$) sets $x_{c0} = 0$. The same argument naturally extends to a more general situation including Type III. Furthermore, the fixed point of $u_q$ sets the quantum vertex to zero, rendering it irrelevant as we discussed in the main text. Finally, the parameter $r$ characterizing the distance from critical point increases under scaling, and should be tuned to zero to access the critical point itself as is typical in the analysis of critical behavior.

The above scaling behavior allows us to write the correlation and response functions in a general scaling form as

$$C(t_2, t_1) = \frac{1}{4}\langle x_c(t_2) x_c(t_1) \rangle = \lambda^{\frac{2(1-\alpha_0)}{1+\alpha_0}} \hat{C}\left(\frac{t_1}{\lambda}, \frac{t_2}{\lambda}, \lambda^2 r, \frac{\lambda^{\frac{4}{1+\alpha_0}}}{N}\right), \tag{S24a}$$

$$\chi^R(t_2, t_1) = \frac{-i}{4}\langle x_c(t_2) x_q(t_1) \rangle = \lambda\, \hat{\chi}^R\left(\frac{t_1}{\lambda}, \frac{t_2}{\lambda}, \lambda^2 r, \frac{\lambda^{\frac{4}{1+\alpha_0}}}{N}\right), \tag{S24b}$$

where the correlation and response functions $C$ and $\chi^R$ should be identified with those in Eqs. (5b) and (5c) of the main text (at all times $t_1, t_2 > 0$); the response function introduced here is related to that in the main text via $\chi^R = \Theta(t_2 - t_1)\chi$ with $\Theta$ the Heaviside step function. We have also introduced the scaling functions $\hat{C}$ and $\hat{\chi}^R$. Note that the scaling parameter $\lambda$ can be chosen as desired. To recover the finite-size scaling in Eqs. (5b) and (5c) of the main text, we use the fact that, at late times, the correlation and response functions should only depend on $t_2 - t_1$; additionally, we set $r = 0$ and $\lambda = N^{\frac{1+\alpha_0}{4}}$. Finally, to characterize the fluctuations in Eq. (5a), we set $t_1 = t_2$ in the correlation function $C(t_2, t_1)$. With these substitutions, we recover the scaling forms in Eqs. (5a) to (5c) with the finite-size scaling exponents

$$\alpha = \frac{1-\alpha_0}{2}, \qquad \zeta = \frac{1+\alpha_0}{4}, \tag{S25}$$

as reported in the main text.

The scaling behavior in the thermodynamic limit ($N \to \infty$) away from the critical point ($r > 0$) can also be identified from the same scaling relations. To this end, we consider the equal-time correlation function at late times when this function has reached a time-independent stationary value. The scaling behavior of fluctuations in the thermodynamic limit $N = \infty$ can be identified by setting $\lambda = 1/\sqrt{r}$ in Eq. (S24a) to find

$$\lim_{t \to \infty} C(t,t) \sim \frac{1}{r^{\frac{1-\alpha_0}{1+\alpha_0}}} = \begin{cases} 1/r, & \text{Type I (Thermal)}, \\ 1/\sqrt{r}, & \text{Type II (Quantum)}, \\ 1/r^2, & \text{Type III (Non-equilibrium)}. \end{cases} \quad (S26)$$

(While our analysis does not apply to Type II, the above equation also correctly reproduces the scaling behavior at the quantum critical point, which is reported for completeness.) The divergence of the fluctuations with the distance from the critical point can be thus identified as a new critical exponent,

$$\nu = \frac{1-\alpha_0}{1+\alpha_0}. \quad (S27)$$

While this exponent takes the known equilibrium (thermal and quantum) values $\nu = 1$ and $\nu = 1/2$ for Type-I and II quenches, respectively, it assumes a different value $\nu = 2$ for the Type-III quench, identifying a genuinely non-equilibrium critical behavior.

### S.IV. EFFECTIVE TEMPERATURE: A SCALING ANALYSIS

In this section, we will identify the effective temperature based on the fluctuation-dissipation relation. To this end, we consider the correlation and response functions at long times when the system has approached a stationary state. In this limit, the two-time correlators only depend on the time difference, $t_2 - t_1$. It will be more convenient to go to the Fourier space, $C(\omega) = \int d(t_2-t_1) e^{i\omega(t_2-t_1)} C(t_2-t_1)$ and similarly for $\chi$. The scaling functions in Eqs. (5b) and (5c) of the main text can be cast in the frequency domain as

$$C(\omega) = N^{\alpha+\zeta} \tilde{C}(\omega N^\zeta), \quad (S28a)$$
$$\chi(\omega) = N^{2\zeta} \tilde{\chi}(\omega N^\zeta), \quad (S28b)$$

where we have set $r = 0$ and, in an abuse of notation, have used the same symbols to denote the scaling functions in frequency space. At low frequencies ($|\omega| N^\zeta \ll 1$ in units of $J$), the scaling functions can be Taylor-expanded as

$$\tilde{C}(x) \sim 1 + \cdots, \quad (S29a)$$
$$\tilde{\chi}(x) \sim ix + \cdots. \quad (S29b)$$

The form of the Taylor expansion follows from the fact that the correlation function is real valued and even in $\omega$ while the response function, characterizing dissipation, is odd in $\omega$, i.e., odd under time reversal symmetry. (In this section, we do not keep track of the precise coefficients as we are only interested in the scaling behavior.) The expansion at small frequencies yields the form of the two-point correlators in frequency space as

$$C(\omega) \sim N^{\alpha+\zeta} + \cdots, \quad (S30a)$$
$$\chi(\omega) \sim iN^{3\zeta}\omega + \cdots. \quad (S30b)$$

The effective temperature at low frequencies can be then identified from the classical limit of the fluctuation-dissipation relation, $C(\omega) = (2T/\omega) i\chi(\omega)$ [S1], as

$$T^{\text{IR}}_{\text{eff}} \sim \lim_{\omega \to 0} \frac{\omega C(\omega)}{i\chi(\omega)} \sim N^{\alpha-2\zeta} = N^{-\alpha_0}. \quad (S31)$$

In the last equality, we have used the form of the critical exponents in Eq. (S25) in terms of $\alpha_0$ as identified in the main text (see the end of Sec. S.II). We can then identify the scaling of the effective temperature in the IR limit as

$$T^{\text{IR}}_{\text{eff}} \sim \begin{cases} \text{finite}, & \text{Type I (Thermal)}, \\ 0, & \text{Type II (Quantum)}, \\ N^{1/3}, & \text{Type III (Non-equilibrium)}. \end{cases} \quad (S32)$$

Here, the behavior for Type II is also reported for completeness.

Next we consider the behavior of the effective temperature at large frequencies $|\omega| N^\zeta \gg 1$; we nevertheless consider the frequency to be small compared to the microscopic energy scales (for example, $J=1$). In this limit, correlation and response functions are not regularized by $N$ and only depend on $\omega$. This is the case provided that the scaling functions describing the two-point functions in Eq. (S28) scale as

$$\tilde{C}(x) \sim \frac{1}{|x|^{(\alpha+\zeta)/\zeta}}, \tag{S33a}$$

$$\tilde{\chi}(x) \sim \frac{i\,\mathrm{sgn}(x)}{x^2}, \tag{S33b}$$

at large frequencies (i.e., large $x$). The (non-analytic) sign function is to ensure that the response function $\chi$ is odd in $\omega$. The large-frequency expansion then yields

$$C(\omega) \sim \frac{1}{|\omega|^{\frac{\alpha}{\zeta}+1}} + \cdots, \tag{S34a}$$

$$\chi(\omega) \sim \frac{i\,\mathrm{sgn}(\omega)}{\omega^2} + \cdots. \tag{S34b}$$

A frequency-dependent effective temperature is then obtained as

$$T_{\mathrm{eff}}(\omega) \sim \frac{\omega C(\omega)}{i\chi(\omega)} \sim |\omega|^{-\frac{\alpha}{\zeta}+2} = |\omega|^{\frac{4\alpha_0}{1+\alpha_0}}, \tag{S35}$$

where, in the last equality, we have cast the exponent in terms of $\alpha_0$ via Eq. (S25). We then find that at large frequencies, $N^{-\zeta} \ll \omega(\ll 1)$ in units of $J$, the effective temperature scales as

$$T_{\mathrm{eff}}(\omega) \sim \begin{cases} \text{finite,} & \text{Type I (Thermal),} \\ 0, & \text{Type II (Quantum),} \\ 1/\omega^2, & \text{Type III (Non-equilibrium).} \end{cases} \tag{S36}$$

Once again, the quantum case is reported for completeness.

Finally, we point out that the effective temperature for the Type-III quench can be written in a scaling form directly from Eq. (S28) as $T_{\mathrm{eff}}(\omega) = N^{1/3} \mathcal{T}(\omega N^{1/6})$ with $\mathcal{T}$ a scaling function. At low frequencies ($\omega \to 0$), the scaling function $\mathcal{T}$ saturates to a constant recovering the $N^{1/3}$ scaling described above. On the other hand, at higher frequencies, things are not sensitive to the IR cutoff (i.e., system size, $N$) and should be independent of $N$ requiring $\mathcal{T}(x) \sim 1/x^2$ and resulting in an effective temeprature $T_{\mathrm{eff}}(\omega) \sim 1/\omega^2$, again as described above. The form of the effective temperature at intermediate frequencies is interpolated via the scaling function $\mathcal{T}$. Notably, the effective temperature diverges with either $N \to \infty$ in the infrared ($\omega = 0$) or $\omega \to 0$ in the thermodynamic limit ($N = \infty$).

---



\end{document}